\documentclass[aps,prl,twocolumn,superscriptaddress,floatfix,amsfonts,citeautoscript]{revtex4}
\usepackage{times}
\usepackage{graphicx}
\usepackage{graphics}
\usepackage{dcolumn}
\usepackage{bm}
\usepackage{color}
\usepackage{amssymb}
\usepackage{float} 
\begin{document}

\title{\textbf{Spatio-temporal correlations in Wigner molecules}}

\author{Biswarup Ash}
\affiliation{Indian Institute of Science Education and Research-Kolkata, Mohanpur, India-741246}

\author{J. Chakrabarti}
\affiliation{S.N. Bose National Centre for Basic Sciences, Block-JD, Sector-III, Salt Lake, Kolkata-700098}

\author{Amit Ghosal}
\affiliation{Indian Institute of Science Education and Research-Kolkata, Mohanpur, India-741246}

\begin{abstract}

The dynamical response of Coulomb-interacting particles in nano-clusters are analyzed at different temperatures characterizing their solid- and liquid-like behavior. Depending on the trap-symmetry, both the spatial and temporal correlations undergo slow, stretched exponential relaxations at long times, arising from spatially correlated motion in string-like paths. Our results indicate that the distinction between the `solid' and `liquid' is soft: While particles in a `solid' flow producing dynamic heterogeneities, motion in `liquid' yields unusually long tail in the distribution of particle-displacements. A phenomenological model captures much of the subtleties of our numerical simulations.

\end{abstract}

\maketitle

Study of nano-systems demands refinement of fundamental concepts, in addition to their technological promises. For example, a sharp transition~\cite{Huang,YL1_52} between the solid and liquid turns into a smooth crossover in nano-clusters~\cite{BP94, GhosalNatPhys06, DA13,Bonitz}. While the relaxation in simple solids and liquids produce phonon and diffusive modes, particles' motion in confinements show anomalous behavior~\cite{glassydyn, Anomaly1, Anomaly2, Anomaly3}. Experimental evidences from dipolar colloids~\cite{Melzer1} and dusty plasma~\cite{Melzer2}, point towards a rich interplay of long-range interactions and confinements yielding subtle dynamics, and has attracted theoretical attention~\cite{BP94,TH1,TH2}. Then, how good are these `solid' and `liquid' in traps from a dynamical perspective? We address these questions focusing on the dynamical properties of Coulomb-interacting particles in confinements -- with and without symmetries.

The `solid' phase of Coulomb-interacting particles in a nano-cluster is called a Wigner molecule (WM) because it mimics the physics of a Wigner crystal~\cite{Wigner}. The WM `melts' into a `liquid' in circular~\cite{BP94} and irregular~\cite{DA13} confinements, termed circular and irregular Wigner molecule (CWM and IWM). A comparison between the two elucidates the role of `disorder', coming from the irregularities in the trap, on the crossover. 

Analysis of melting using spatio-temporal correlations yields crucial dynamical information~\cite{Zahn00}. The interplay of interactions and Brownian motion (consistent with the trap-geometry) often leads to frustrated motion and dynamic heterogeneities (DH)~\cite{DH_Rev}, i.e. multiple time-scale relaxations. These are ubiquitous near the transitions in colloids~\cite{Kim13}, gels~\cite{Gao07}, glasses~\cite{Pinaki07,Kob07}, and bio-molecule suspensions~\cite{wang09}.

We consider simple models for IWM and CWM and analyze the equilibrium motion of the constituents using iso-kinetic molecular dynamics (MD) simulations~\cite{FrenkelBook}. Our main results can be summarized as follows:

\noindent (1) The motion in traps display a variety of non-Gaussian behavior of the distribution of particle displacements, including its stretched-exponential {\it spatial} decay. Such an ultra-slow relaxation is consistent with preliminary experimental signatures~\cite{He13}.\\
\noindent (2) The dynamics at long $t$ is intriguing: At low $T$ the solid `flows' producing DH, whose nature depends on the confinement. Though particle displacements are random at large $T$, their distribution defies central limit theorem~\cite{CLT}.\\
\noindent (3) Our results demonstrate how the innate dynamics sets apart the manifestation of Wigner physics~\cite{Wigner} based on the trap symmetries.

We model a WM using the Hamiltonian: ${\cal H} = \phi \sum_{i<j} |\vec{r}_i - \vec{r}_j |^{-1} + \sum_i V_{\rm conf}(\vec{r}_i)$, where the first term represents Coulomb repulsion between particles of charge $q$ and $\phi=q^2/\epsilon$, and $\epsilon$ is the dielectric constant. Here, $V_{\rm conf}$ is the confinement potential. We compare the results from two confinements: (a) Irregular $V_{\rm conf}^{\rm Ir}(\vec{r})=a\{ x^4/b+by^4-2\lambda {x^2}{y^2} +\gamma (x-y)xyr\}$~\cite{Bohigas93, DA13}, and (b) Circular $V_{\rm conf}^{\rm Cr}(\vec{r})= \alpha r^2$, where $\alpha = m \omega_0^2/2$. 

\begin{figure}[t]
\includegraphics[width=8.5cm,keepaspectratio]{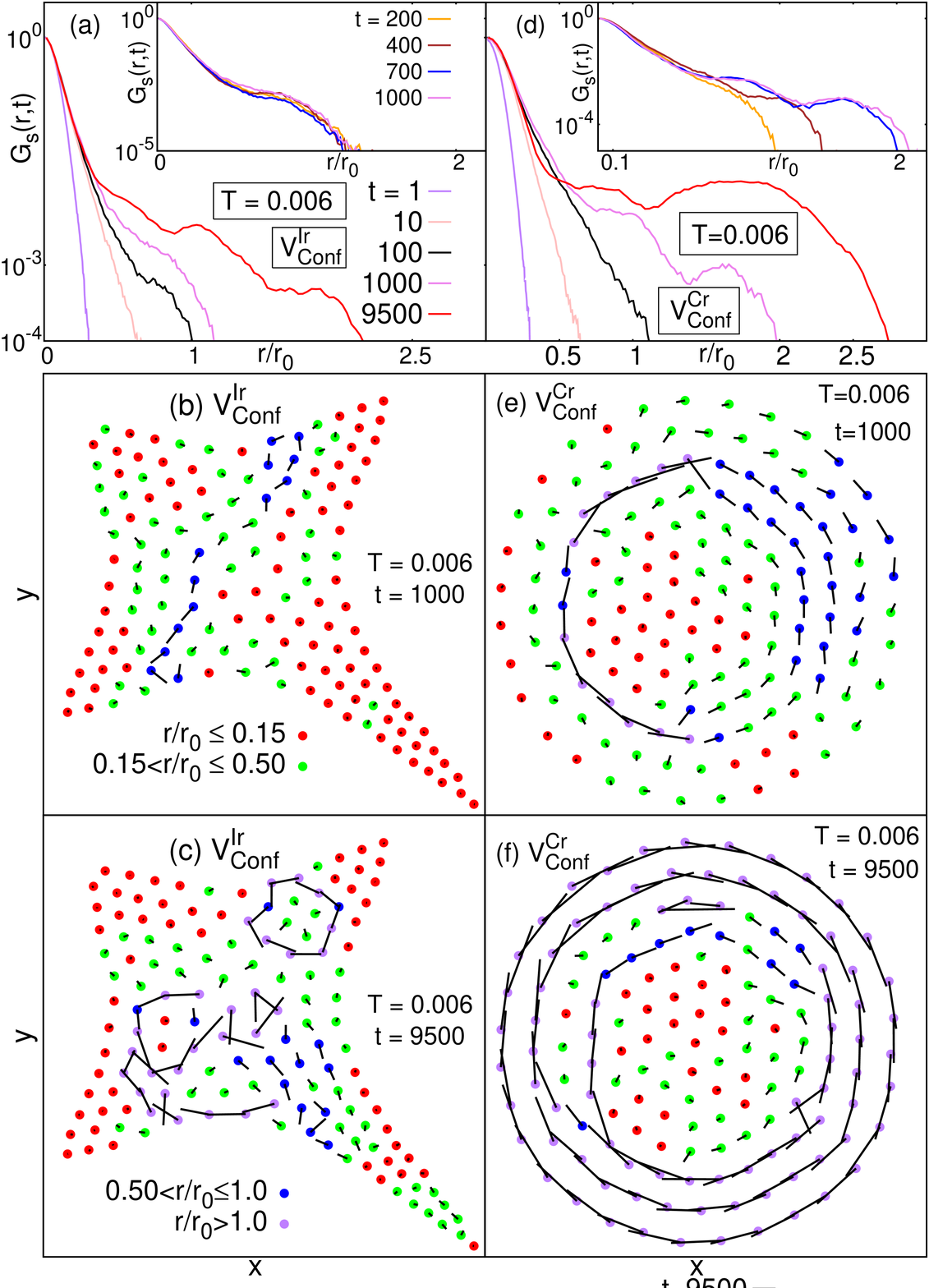} 
\caption{
The decay of $G_s(r,t)$, normalized by $G_s(r=0,t)$ for clarity, for several $t$ in the `solid' ($T=0.006$): (a) IWM, and (d) CWM. $G_s(r)$ is Gaussian up to $t \sim 1$ and displays non-Gaussian trends for $t\geq 10$, including multi-peaks at large times. Bunching-up of the traces of $G_s(r)$ for $100 \leq t \leq 1000$ is shown in the insets demonstrating caging (weaker in CWM).
Corresponding $\Delta \vec{r}(t)$ plots are shown (b, c: IWM and e, f: CWM) for $t=1000$ and $t=9500$, illustrating the motional coherence. The thick dots represent the initial position and the connecting lines signify the displacements of the same particle at $t$. Red, green, blue, and purple dots represent particles with increasing $|\Delta \vec{r}(t)|$.Mobile particles reside in the central region for $V_{\rm conf}^{\rm Ir}$, leaving nearly immobile ones in the narrow limbs near corners. In contrast, the mobile particles in CWM circle around the periphery, while the slow ones flock towards the origin.
}
\label{fig:Self_VH}
\end{figure}

We rescale the length ($r \rightarrow \phi^{1/3}\alpha^{-1/3} r$) and energy ($E \rightarrow \phi^{2/3}\alpha^{1/3} E$), that transform~\cite{BP94} the CWM Hamiltonian to: ${\cal H}^{\rm Cr}=\sum_i r_i^2 + \sum_{i<j} |\vec{r}_i-\vec{r}_j|^{-1}$. The time-scale is set by $ t \rightarrow \hbar~\phi^{-2/3}\alpha^{-1/3} t$~\cite{FN0}. The overall factor $a$ and $\alpha$ are chosen to keep mean density in the two traps same for a fixed total number of particles, $N=150$. The irregularity parameters $b=\pi/4$, $\lambda \in [0.565,0.635]$ (controls chaoticity), and $\gamma \in  [0.10,0.20]$ (breaks reflection symmetry)~\cite{Hong03} are adjusted to access the standard features of disordered systems. Distances are measured in the unit of $r_0$, the mean inter-particle spacing between the neighbors. The statistics of the results from $V_{\rm conf}^{\rm Ir}$ were improved by `disorder averaging'~\cite{Ullmo03} over five configurations, each identified by a specific $\{\lambda,\gamma\}$. For $V_{\rm conf}^{\rm Cr}$, averaging is done over five independent MD simulations.

\begin{figure}[t]
\includegraphics[width=8.5cm,keepaspectratio]{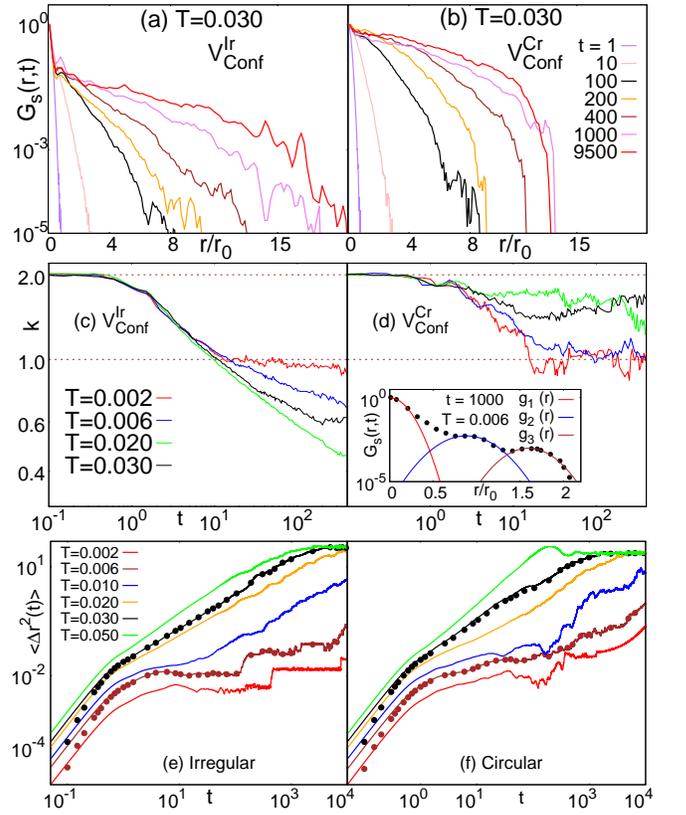} 
\caption{
The evolution of $G_s(r,t)$ in the `liquid' ($T=0.03$) of an (a) IWM and (b) CWM, shows deviations from the Gaussian behavior (except for small $t$) in both traps though $\Delta \vec{r}(t)$ remains random. The nature of the evolution of $G_s(r,t)$ is analyzed in (c) and (d) by studying the exponent $k(t)$ (see text). $k$ decays continuously from $2$ to nearly $1$ at low $T$. The circular trap yields $2 \leq k \leq 1$. $k$ keeps degrading with $t$ and $T$ in irregular confinements. The inset in (d) shows the Gaussian fit of all the three peaks of a representative $G_s(r,t)$ in CWM. $t$-dependence of MSD at different temperatures are shown for irregular (e), and circular (f) traps. The dots represent MSD by integrating $r^2G_s(r,t)$ (see text), and are shown for $T=0.006, 0.03$. The perfect match between the solid lines and corresponding dots persists for all $T$.
}
\label{fig:Snap}
\end{figure}

A thermal crossover from a `solid' to `liquid' in WM asserts~\cite{DA13,BP94} a `solid' at $T=0.006$ and `liquid' at $T=0.03$ for our chosen parameters. The spatio-temporal evolution of the system is studied from the Van Hove correlation function (VHCF)~\cite{VH54}, whose self-part is~\cite{HansenBook06}: $G_s(r,t)= \langle N^{-1}\sum^{N}_{i=1}\delta \left[ r - |\vec{r}_i(t) - \vec{r}_i(0)|\right] \rangle$. Here $\langle .\rangle$ denotes average over up to $5000$ independent time origins. $G_s(r,t)$ thus measures the probability of a particle to move (on average) a distance $r$ in time $t$. Focusing on IWM at $T=0.006$ in Fig. 1(a), the dynamics can be broadly classified into three temporal regimes~\cite{FN1}: (i) The particles move ballistically for $t \leq 1$, and the resulting $G_s(r,t)$ is Gaussian. They start feeling each other only for $t>1$ implying that our unit time is the time scale of interaction. (ii) Particles get `arrested' in the cage formed by neighbors in the intermediate time $(100 \leq t \leq 1000)$, leading to a slowly decaying $G_s(r,t)$ whose traces bunch up, see the inset of Fig. 1(a). (iii) The dynamics for $t \geq 1000$ develop multi-peak structure in $G_s(r,t)$.

In order to develop a deeper understanding, we next analyze the particle displacements, $\Delta \vec{r}_i(t) = \{\vec{r}_i(t) - \vec{r}_i(0)\}$. Most particles only rattles around their equilibrium positions for $t \ll 1000$. Motion that begins by breaking a cage causes avalanches of such events correlated in space, which leads to a string-like path of delocalization~\cite{Glotzer,Anomaly1}. Fig.~1(b) presents $\Delta \vec{r}_i(t=1000)$ for an IWM at $T=0.006$, showing such an alignment of neighboring particles leading to a slow relaxation that yields a long tail in $G_s(r,t)$. Once the collective motion subsumes large part of the system, particles undergo little relative displacements, but an overall drift, producing secondary peak in $G_s(r,t)$. With further increase in $t$, the string-like path degenerates into sub-clusters, each performing independent collective motion (Fig.~1(c)), and contribute to the multi-peak $G_s(r,t)$ of Fig.~1(a).

The motion in the CWM at $T=0.006$ in Fig.~1(d) is qualitatively similar to IWM insofar as the temporal regimes are concerned, albeit the caging regime is narrow ($10 \leq t \leq 100$) and untenable, causing weaker agglomeration of $G_s(r,t)$ (inset, Fig.~1(d)). However, the differences between the two traps stand out in the nature of coherent motion. While the circular symmetry forces such motion along its azimuth (Fig.~1(e, f)), the collective dynamics in IWM is guided by the irregularity producing nonequivalent regions (Fig.~1(b, c)). Unlike the sub-clusters with independent motions in IWM, a single (azimuthal) coherence grows in the CWM, see Fig.~1(f). This results into a rather broad hump of $G_s(r,t=9500)$ in Fig.~1(d). Interestingly, the mobile particles travel by several lattice spacings riding on the coherence even in this `solid' phase~\cite{FN2}. Signatures of spatially correlated motion of particles, have been realized in circular plasma crystals~\cite{Melzer2}, and also in a flocking transition of a driven granular matter in confinements~\cite{Sood_Swamy}. 

In the `liquid' ($T=0.03$), $\Delta \vec{r}(t)$ are large and random. Yet, $G_s(r,t)$ deviates from Gaussian and differences persist in the two traps (Fig.~2(a, b)). While the displacements are nearly Gaussian in circular traps, it undergoes a progressively slower decline for $V_{\rm conf}^{\rm Ir}$. It is not possible to discern sharp temporal regimes in such a `melted' state.

Next, we quantify the large-$r$ decay of $G_s(r,t)$ by fitting it to $\sim e^{-l r^k}$ (See supplementary material). Fig.~2 (a: IWM, b: CWM) presents the time-evolution of the best fit parameter $k$. For low $T$, $k$ degrades close to unity (both traps) establishing a near-exponential decay for intermediate times. With an increase in $T$, the long-time value of $k$ rises in $V_{\rm conf}^{\rm Cr}$, however, broadly remains within $1 \leq k \leq 2$. In a striking contrast, $k$ decreases monotonically for $V_{\rm conf}^{\rm Ir}$, causing a {\it stretched} exponential spatial decay.

Above analysis does not extend beyond $t = 400$ due to the non-monotonic evolution of $G_s(r,t)$ at larger $t$ and low $T$. The monotonic fall of $k(t)$ in the irregular trap, however, continues for large $T$, and hints towards a saturation only for $t \geq 8500$. Interestingly, $k$ reaches its minimum for $T=0.02$, a temperature that marks the crossover between the `solid' and `liquid'~\cite{DA13}. The traces of $k(t)$ rises for $T>0.02$. The system resembles a conventional liquid ($k \approx 2$ for large $t$) only at a much larger $T\sim 0.25$. In contrast, a circular confinement yields $k \rightarrow 2$ at large $t$ for $T \geq 0.05$ (See supplementary material).

The classification of temporal regimes is also revealed from the mean-square displacement (MSD), $\langle\Delta r^2 (t) \rangle$, shown in Fig. 2 (e: IWM, f: CWM). The low-$T$ evolution of MSD is surprisingly flat for IWM showing strong caging. The corresponding CWM show a weak Brownian behavior. The plateau at intermediate $t$ gradually shrinks with $T$ leading to a smooth crossover between ballistic and Fickian motion. It follows that, $\langle \Delta r^2 (t)\rangle = \int r^2 G_s(r,t) d^2r$, and its estimation is shown as dots in Fig. 2(e-f), confirming the `Fickian yet non-Gaussian' motion~\cite{Kim13}  in the WM at large $t$. MSD saturates when displacements approach the system size. Its occasional jumps in IWM fades away with $T$. Such jumps are not discernible in a CWM even at the lowest $T$. IWM, thus, resembles a supercooled system~\cite{DH_Rev} more than what CWM does. 

\begin{figure}[t]
\includegraphics[width=8.5cm,keepaspectratio]{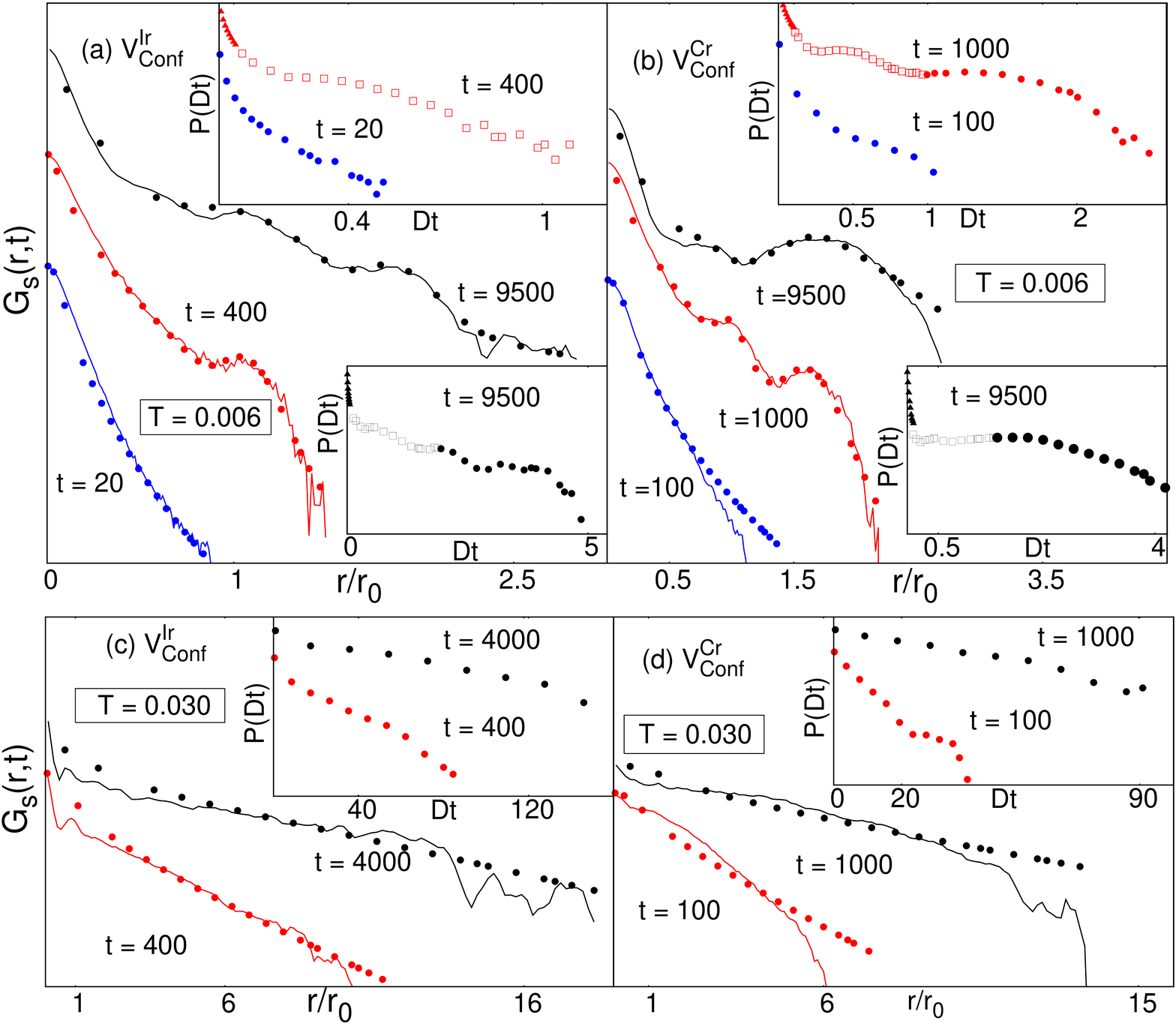}
\caption{
The main panels show the comparison of $G_s(r,t)$ obtained from direct MD simulations with its evaluation using Eq. (1). The insets show corresponding $P(D)$ and the parameters $t$ and $T$ appear as legends. The nature of $G_s(r,t)$ appears largely controlled by  $P(D)$, from its rich structures due to heterogeneous diffusion, and from its long tail.
}
\label{fig:g6Gd}
\end{figure}

A phenomenological description expresses $G_s(r,t)$ as a convolution of Gaussian displacements and the distribution of diffusivities, 
$P(D)$~\cite{Th1,wang12}. We generalize this to accommodate the multi-peak nature of $G_s(r,t)$:
\begin{equation} \label{eq:suposeq}
G_s(r,t) = \sum_{p=1}^{N_p} \int dD_p \frac{P(D_p)}{(4\pi D_p t)}{\rm exp}\left[-\frac{(r-r_p)^2}{4D_p t}\right],
\end{equation}
where $N_p$ is the number of peaks observed in  $G_s(r,t)$. By fitting the individual peaks (see the inset of Fig. 2(d)), we obtain $r_p$ and the peak widths. We determine $P(D_p)$ from the statistics of particles whose $\langle {r_i}^2(t)\rangle$ falls within the peak width corresponding to $r_p$~\cite{FN3}. A comparison of Eq.~(1) with the direct simulation is shown in Fig. 3(a-d). The distinct $r_p$ signals dynamic heterogeneity for $t \geq 1000$.  The inset of Fig.~3(c) shows a stretched-exponential fall~\cite{Th1} of $P(D)$  for $T=0.03$ in $V^{\rm Ir}_{\rm conf}(r)$, whereas, its near exponential decay characterizes $V^{\rm Cr}_{\rm conf}(r)$. A full $P(D)$ corresponding to a multi-peak $G_s(r,t)$ displays rich structures (insets of Fig.~3(c-f)), and the higher peaks contribute near homogeneous weight to $P(D)$.

\begin{figure}[t]
\includegraphics[width=8.5cm,keepaspectratio]{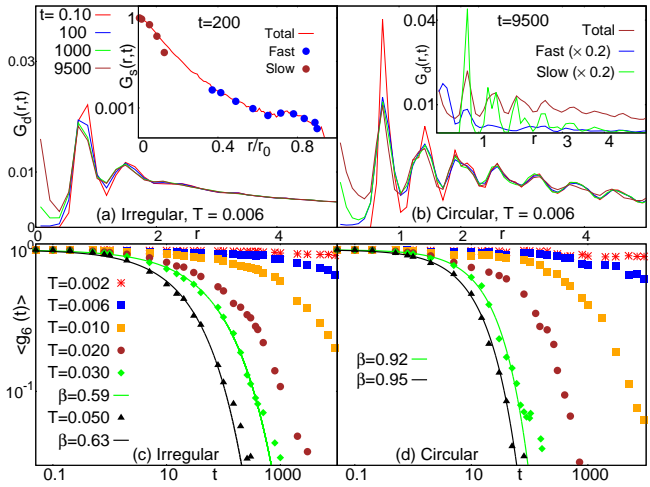}
\caption{
{\it Upper panels:} $G_d(r,t)$ with $r$ for several $t$ in the `solid', depicting the temporal relaxation in (a) IWM and (b) CWM. The `correlation-hole' in $G_d(r \approx 0,t)$ fills up with time producing strong peak at large $t$. The inset of (a) identifies the contribution of slow (brown) and fast (blue) particles in $G_s(r,t)$. The inset of (b) shows that the fast particles contribute to $G_d(r \rightarrow 0,t)$, while the slow ones trace the frozen `solid'. {\it Lower panels:} A strong bond-orientation at $T=0.006$ reflects in $\langle g_6(t) \rangle \approx 1$, for all $t$ for both IWM (panel c) and IWM (panel d). Its decay in the melted state ($T=0.03$) follow an exponential and stretched-exponential $t$-dependences for circular and irregular confinements respectively.
}
\label{fig:DVH}
\end{figure}

The multiple relaxations is also probed by the distinct-part of VHCF, $G_d(r,t)$~\cite{HansenBook06}: $G_d(r,t)=\langle 2[N(N-1)]^{-1}\sum_{i\neq j}\delta \left[ r - |\vec{r}_{i}(t) - \vec{r}_{j}(0)|\right] \rangle$. It represents the probability of finding at $t$ a different particle at a distance $r$ from a location where there was a particle initially. Fig.~4(a) shows the lack of positional order at $T=0.006$ in the IWM~\cite{DA13} through depleted Bragg-peaks~\cite{FN4} of $G_d(r,t\rightarrow 0)$, while a CWM displays sharper peaks (Fig.~4(b)) due to the azimuthal order.

We substantiate the heterogeneity defining mobility~\cite{Kim13} as $\langle |\Delta \vec{r}_i(t)| \rangle$. A particle is qualified as `fast' (`slow') if its mobility lies within the top (bottom)  $7\%$ ($10\%$). The long-time accumulation of $G_d(r \rightarrow 0)$ (Fig.~4(a,b)) arises from the fast particles occupying the location from where another particle moved out, developing spatially correlated motion. In contrast, the slow particles trace out Bragg-peaks. Such distinct behavior is illustrated as the inset of Fig.~4(b) for $V_{\rm conf}^{\rm Cr}$. A similar picture emerges from the self-part too -- while the slow particles produce the small-$r$ Gaussian behavior, the fast particles contribute to the long tail of $G_s(r,t)$, see inset of Fig.4(a).

The `solid' state of a WM possesses bond-orientational order. Each particle has predominantly six neighbors, and each of these when connected to the given particle makes an angle of $\pi/3$ with the line connecting successive neighbors. The corresponding bond-orientational correlation function: $g_{6}(t)= \langle {\rm exp}[i6\theta(t)]\rangle$~\cite{NelsonBook02} is unity for a perfect Wigner crystal. Its decay describes how the propensity of the six co-ordinated neighborhood corrodes in time. Here $\theta(t)$ is the angle fluctuation of a fixed bond. We present $g_6(t)$ in Fig. 4(c) and (d) for $V_{\rm conf}^{\rm Ir}$ and $V_{\rm conf}^{\rm Cr}$ respectively. A bulk 2D isotropic liquid shows an exponential fall of  $g_6(t)$ with $t$ signaling the loss of this order according to the KTHNY theory~\cite{NelsonBook02}. While our results in Fig.~4(d) broadly capture such a trend in $V_{\rm conf}^{\rm Cr}$, the `liquid' in $V_{\rm conf}^{\rm Ir}$ (Fig.~4(c)) shows a slower, stretched-exponential decay: $g_6(t) \sim {\rm exp}\left[-{\alpha} t^{\beta}\right]$ with $\beta \sim 0.6$.

In conclusion, we find non-Gaussian distribution of the particle displacements, correlated motion of the constituents in string-like paths, distinct relaxation of slow and fast particles, and stretched-exponential decay of both spatial and temporal correlations in irregular confinement with Coulomb particles, whereas the circular traps produce different motional signatures. Bulk supercooled liquids, which are also disordered systems with short-range inter-particle interactions, share much of these dynamic footprints. Such similarities hint towards these findings to constitute the robust dynamical features of disordered systems. It will be interesting to explore the response functions~\cite{SmarajitRev} in different temporal regimes. Our results will provide guidance in designing nano-technology devices.
 
{\it Acknowledgment}- We thank Deepak Dhar and Chandan Dasgupta for valuable discussions. BA acknowledges his doctoral fellowship from the University Grant Commission (UGC), India.

%%%%%%%%%% Merge with supplemental materials %%%%%%%%%%
\pagebreak
\widetext
\clearpage % added by PV
~\vspace{2cm} % added by JG

%%%%%%%%%% Merge with supplemental materials %%%%%%%%%%
%%%%%%%%%% Prefix a "S" to all equations, figures, tables and reset the counter %%%%%%%%%%

\setcounter{equation}{0}
\setcounter{figure}{0}
\setcounter{table}{0}
\setcounter{page}{1}
\makeatletter
\renewcommand{\theequation}{S\arabic{equation}}
\renewcommand{\thefigure}{S\arabic{figure}}
\renewcommand{\bibnumfmt}[1]{[S#1]}
\renewcommand{\citenumfont}[1]{S#1} % might need to play with \renew v \new
%%%%%%%%%% Prefix a "S" to all equations, figures, tables and reset the counter %%%%%%%%%%

\begin{center}
{\Large\bf Spatio-temporal correlations in Wigner molecules: Supplementary Information}
\end{center}

\vspace{0.5cm}

\begin{center}
\vspace{0.5cm}

Biswarup Ash\textsuperscript{1}, J. Chakrabarti\textsuperscript{2}, and Amit Ghosal\textsuperscript{1}

\vspace{0.5cm}
{\em \textsuperscript{1}Indian Institute of Science Education and Research-Kolkata, Mohanpur, India-741246.

     \vspace{0.3cm}
     \textsuperscript{2}S.N. Bose National Centre for Basic Sciences, Block-JD, Sector-III, Salt Lake, Kolkata-700098.}\\[15mm]

\end{center}

We begin this supplementary material elaborating the procedure of obtaining $k$ as a function of $t$, as presented in Fig.~2(c, d) of the main paper. 

\section{Extraction of the exponent $k(t)$}

\begin{figure}[H]
\centering
      \includegraphics[width=9cm,keepaspectratio]{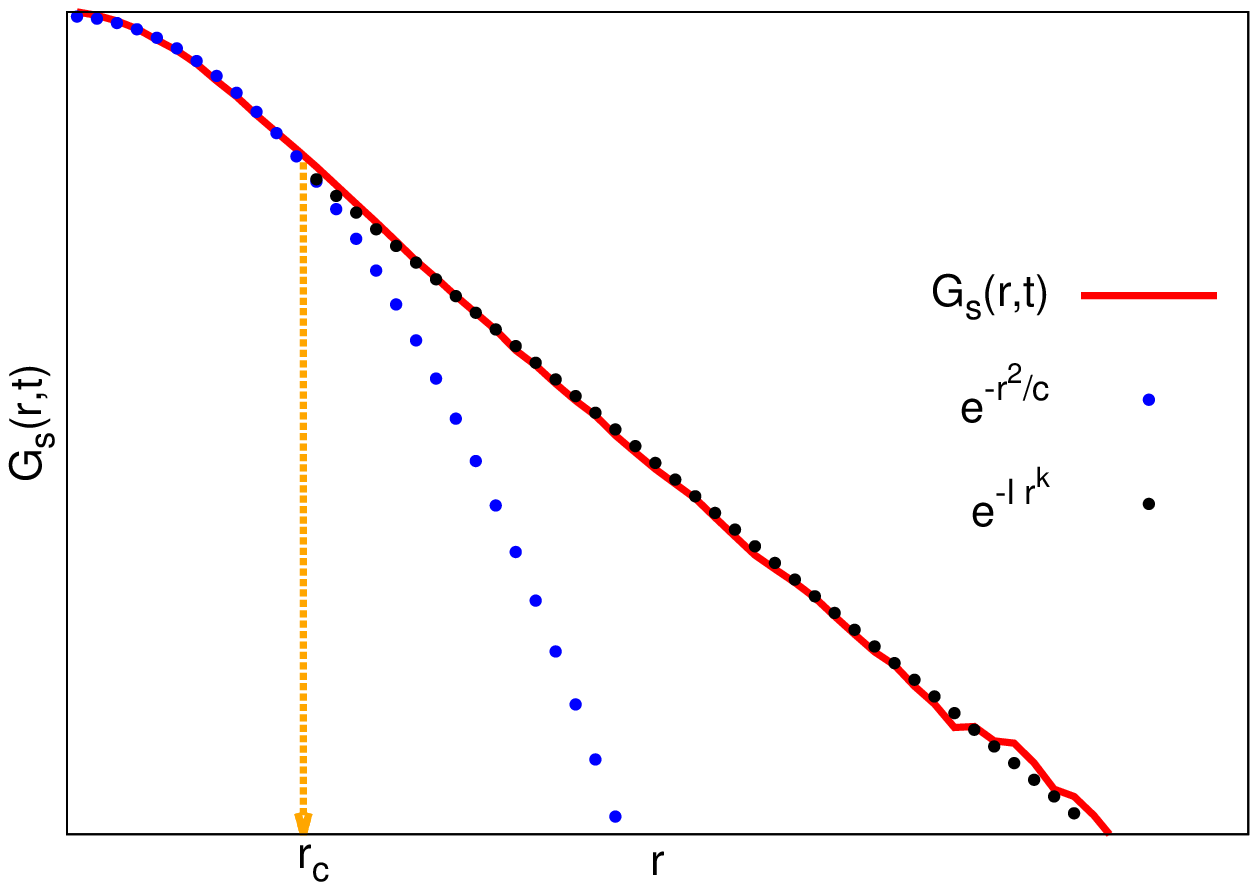} 
      \caption{Illustration of the fitting of $G_s(r,t)$ curve with two functional dependencies: $G_{s}^{\rm small}(r,t) \sim e^{-{r^2}/c}$  for $r \leq r_c $, and $G_{s}^{\rm large}(r,t) \sim e^{-l r^k}$ for $r>r_c $. $r_c$ denotes the optimal distance for which total $\chi^2$ is minimum. }
      \label{fig:Fit_Gs}
\end{figure}

We fit the MD data of $G_s(r,t)$ with two functional dependencies:
$G_{s}^{\rm small}(r,t) \sim e^{-{r^2}/c} ~\text{for}~ r \leq r_c,$ and $G_{s}^{\rm large}(r,t) \sim e^{-l r^k}  ~\text{for}~ r>r_c$
for a range of $\{t,T\}$, for which the $r$-dependence of $G_s(r,t)$ has a monotonic evolution.
%, such that, $G_{s}^{\rm small}$ and $G_{s}^{\rm large}$ make sense. 
For a given $t$ and $T$, we start our fitting procedure by choosing a rather small value of $r_c$, and fit the entire $G_s(r,t)$ curve with either of $G_{s}^{\rm small}(r,t)$ or $G_{s}^{\rm large}(r,t)$ depending on $r$ smaller or larger than the chosen $r_c$. For each region ($r$ smaller or larger than $r_c$), we get a value for $\chi^2$, measuring the goodness of the fit~\cite{Numeric_Recipe} in that region. Adding its values from the two regions, we get the total $\chi^2$ of the entire fit for that specific $r_c$. Now, we increase $r_c$ in small incremental steps, and calculate total $\chi^2$ for each such $r_c$. Finally, the optimal values of $r_c$ and other parameters, e.g., $c, l, k$ were obtained by considering the $r_c$ for which total $\chi^2$ is minimum. The data in Fig.~2(c, d) of the main paper are generated by repeating above procedure for the entire range of $\{t,T\}$. The adjoining Fig.~\ref{fig:Fit_Gs} depicts one such case, where we highlight the above procedure.

We emphasize that a value of $k=2$ implies that the whole $G_s(r,t)$ curve has a single Gaussian fall. While it is also possible to get $k=2$ from a multiple-Gaussian~\cite{Double_Gauss} fall with different width in different ranges of $r$, our results of $k=2$ do not correspond to such possibilities.

\section{Exponent $k$ in the limit of large $t$ and $T$}

In the main text, we have presented the evolution of $k$ with data limited to $t \leq 400$, and $T \leq 0.03$. While extrapolation of the same analysis is not possible at large $t$ for the lowest temperatures, we here present $k(t)$ up to $t \sim 10000$ for $0.02 \leq T \leq 0.1$. The two main features to note here are: \\
(a) For $V_{\rm conf}^{\rm Ir}$, the traces of $k(t)$ rises with $T$ for $T \geq 0.02$, where as initial increase of $T$ for $0.002$ to $0.020$ lowered them progressively (Fig. 2(a) in the main text). Interestingly, $T=0.020 $ marks the solid to liquid crossover according to static properties \cite{DA15}, for which $k$ attains the minimum value (over all $T$).

(b)For $V_{\rm conf}^{\rm Ir}$, the monotonic fall of $k(t)$ for all $T$ persists, and shows a tendency toward saturation for large $t (\geq 8500)$.

\begin{figure}[h!]
\centering
      \includegraphics[width=18cm,keepaspectratio]{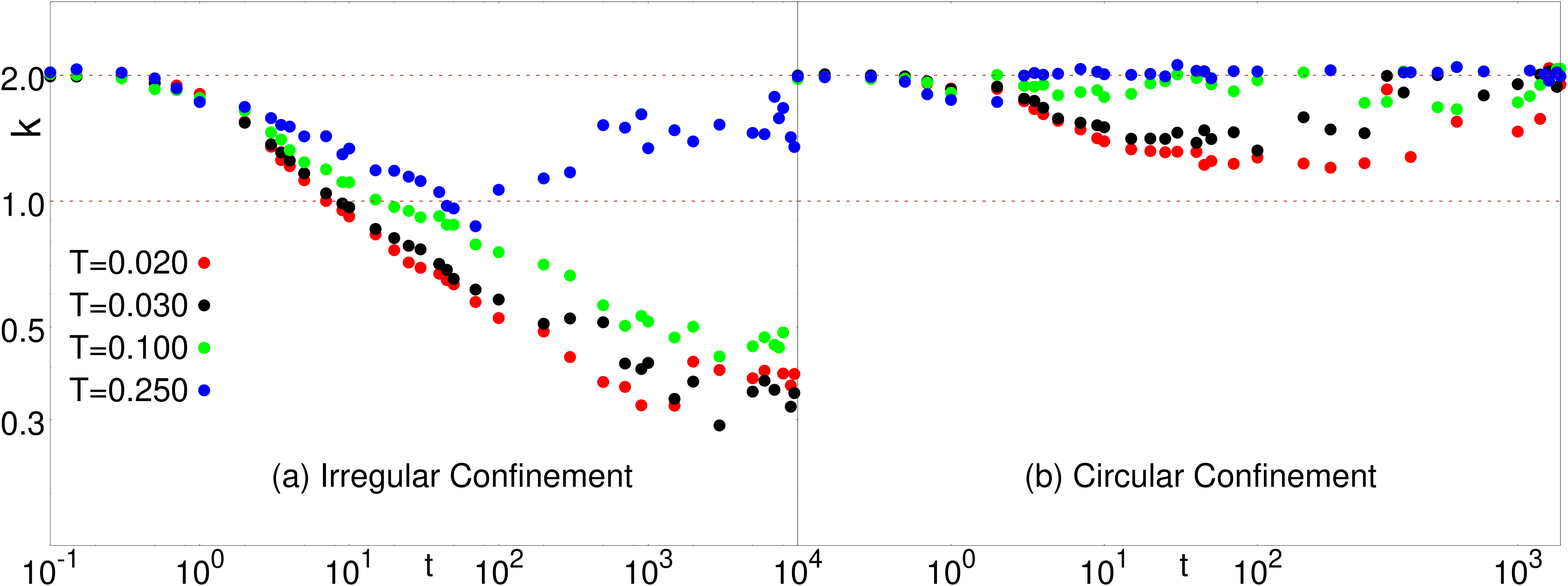} 
      \caption{Evolution of the exponent $k$ in the limit of large time $(t)$ and temperature $(T)$ for the Irregular (panel a) and Circular (panel b) confinements. While, for circular confinement, $k(t)$ is mostly bounded between the values $1$ and $2$, it goes below $1$ and shows a tendency towards saturation for iregular confinement for $T \geq 0.02$.}
      \label{fig:Expk}
\end{figure}

For $V_{\rm conf}^{\rm Cr}$, on the other hand, $k$ approaches a value of $2$, indicating Gaussian behavior for large $t$, and $T$. The data presented in Fig.~\ref{fig:Expk} has results from a single molecular dynamics simulation resulting into large fluctuations, though we made sure that they represent the true situation.

\section{Illustration of caging effects from displacement plots}

\begin{figure}[H]
\begin{center}
      \includegraphics[width=15cm,keepaspectratio]{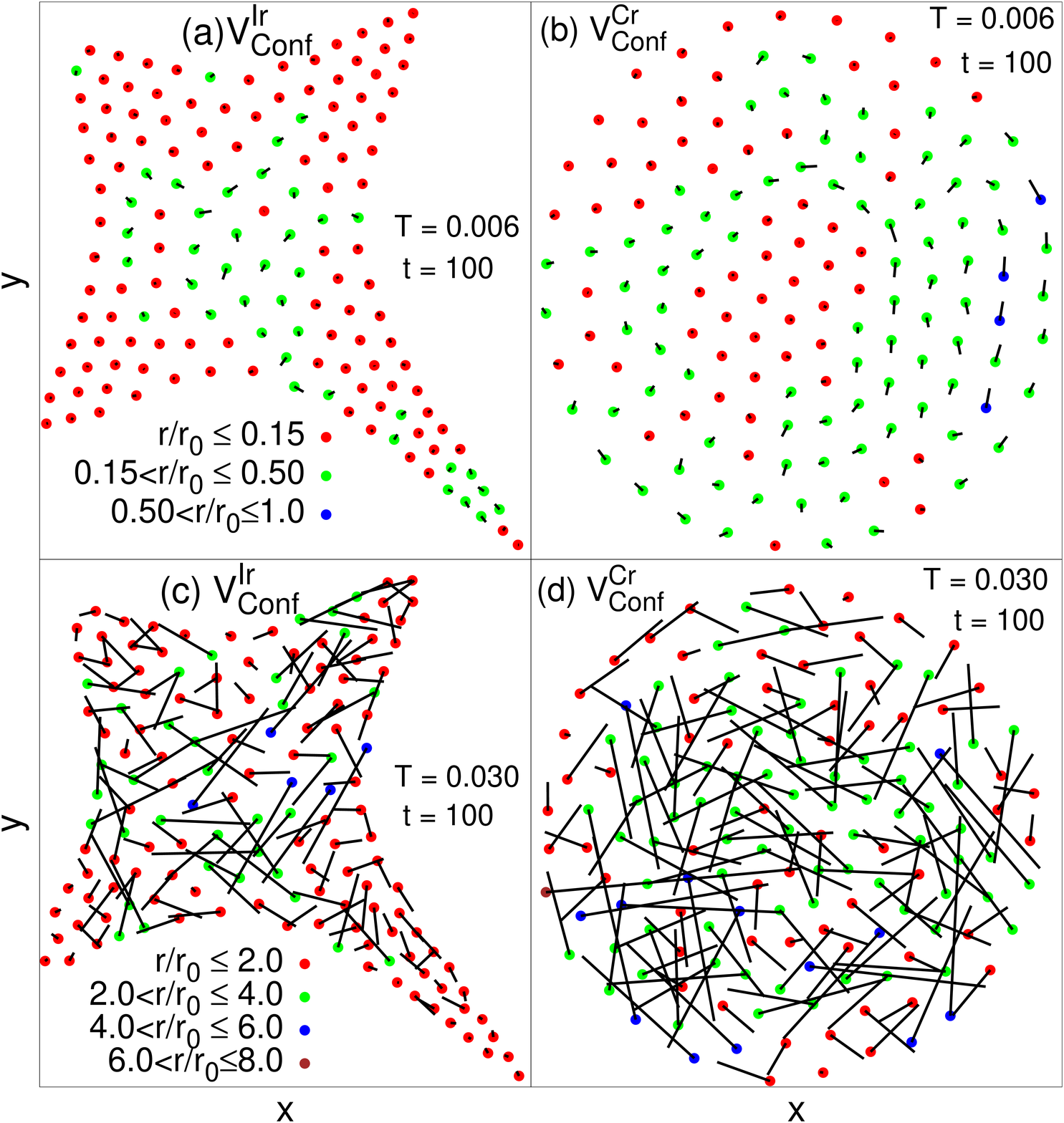} 
      \caption{The displacement $\Delta \vec{r}(t)$ of the particles are shown at $T=0.006$ (panel (a), and (b)), and $T=0.030$ (panel (c), and (d)) for IWM and CWM at $t=100$. Red, green, blue, and brown dots represent initial positions of the particles with increasing magnitude of displacement. The length of the connecting black lines correspond to the magnitude of the displacements of the particles at the given time $t$.
}
      \label{fig:Disp_100}
\end{center}
   \end{figure}

Fig.~\ref{fig:Disp_100}(a, b) shows the displacements $\Delta \vec{r}(t) = \{\vec{r}(t) - \vec{r}(0)\}$ of the particles for irregular and circular confinements, respectively, at intermediate time $(t=100)$ for $T=0.006$, illustrating the caging effect.
For both confinements, most particles (red dots) perform rattling motion around their equilibrium positions. These particles are `trapped' within the cage formed by their nearest neighbors, and such an arrest causes $G_s(r,t)$ traces to collapse (inset of Fig. 1(a) in the main text) on each other for intermediate times. From the alignment of dots and the connecting lines (Fig.~\ref{fig:Disp_100}(a)), we see that the coherent motion sets up randomly due to the absence of spatial symmetry for IWM, while such commencement of the coherent motion in CWM (Fig.~\ref{fig:Disp_100}(b)) is guided by the circular symmetry.\\
Fig.~\ref{fig:Disp_100}(c, d) shows the displacements of the particles for the two confinements in the same time regime for $T=0.030$. Here, we see that the displacements are random and relatively large in magnitude, for both the confinements. One can expect such random motion at this temperature because of the higher thermal energy.

\section{Comparison of  $G_s(r,t)$ obtained from MD simulation, and from evaluation of $Eq. (1)$ in the main text}

\begin{figure}[h]
\begin{center}
      \includegraphics[width=14cm,keepaspectratio]{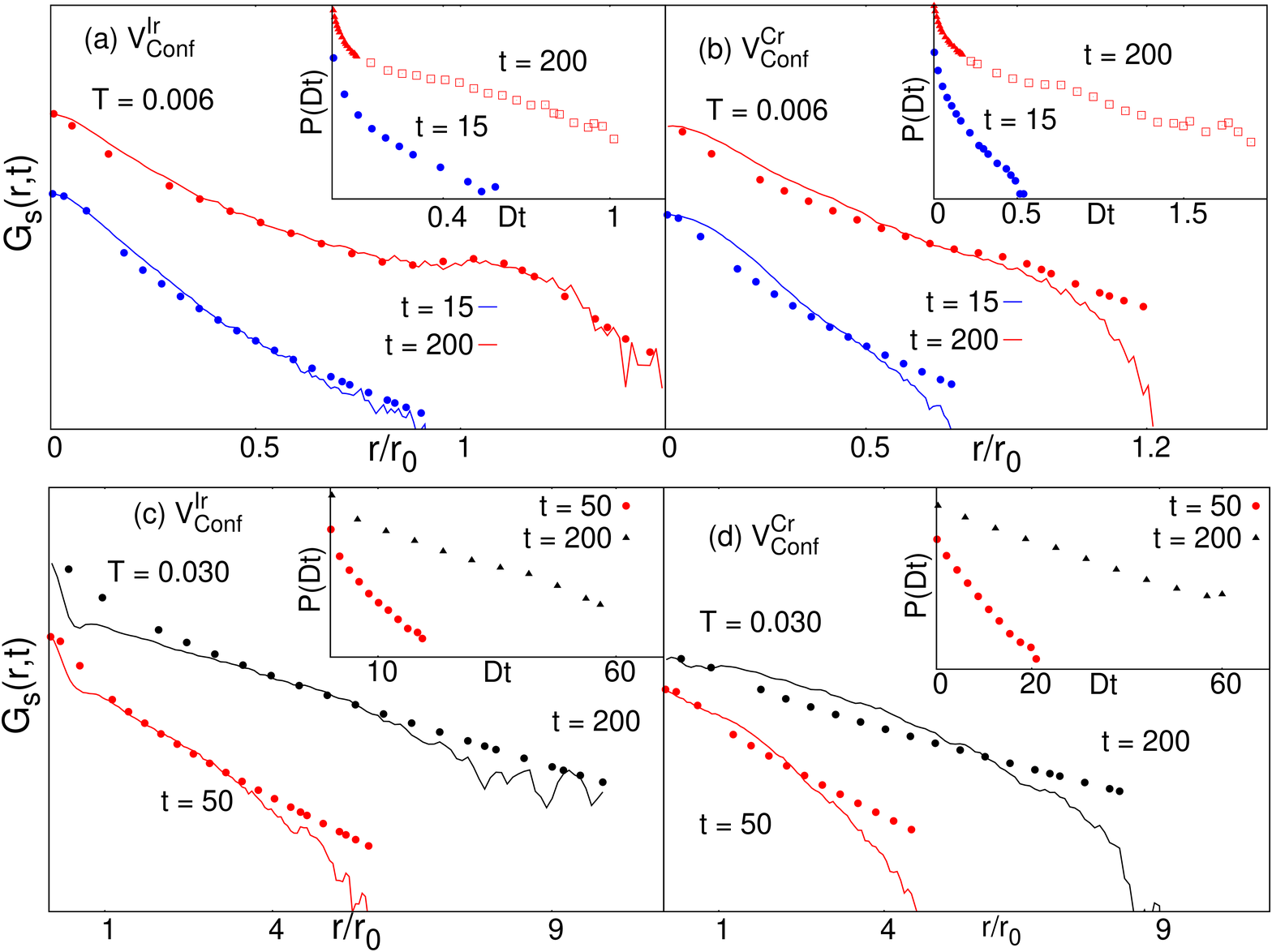} 
      \caption{Comparison of $G_s(r,t)$ from direct MD simulations with its evaluation from Eq. $(1)$ (see main paper) for IWM (Panel (a),(c)) and CWM (Panel (b),(d)) at T=0.006 and 0.030, respectively. Thick dots represent $G_s(r,t)$ evaluated from Eq. $(1)$, and solid lines are the same evaluated from MD simulation. Insets show the distribution of diffusivities $P(D)$ used to evaluate $G_s(r,t)$ for the given $t$.} 
      \label{fig:PD}
\end{center}
   \end{figure}

In our paper, we claimed (Fig. 2 (c-f)) that $G_s(r,t)$ evaluated from Eq. (1) shows broad agreement with its direct calculation from the MD simulation. Here, in Fig.~\ref{fig:PD}, we quantify such `broad' agreement further, for both the confinements. 
At $T=0.006$, while we have already seen $G_s(r,t)$ evaluated from MD simulation shows good match with its evaluation from Eq. (1) for all time regimes for IWM (Fig. 2 (c) in the main paper), CWM shows small deviation at intermediate times ($t=200$, in Fig. 2 (d) in the main paper). Here, we substantiate this point in Fig.~\ref{fig:PD} (a), and (b) for IWM and CWM, respectively. At intermediate times $(t=15,200)$, we see good agreement for IWM (Fig.~\ref{fig:PD} (a)), while small deviation persists for CWM (Fig.~\ref{fig:PD} (b)).
At $T=0.030$, we found that our phenomenological formalism could not capture the large $r$ trend of the $G_s(r,t)$ curve for CWM (Fig. 2 (f) in the main paper). Here, from Fig.~\ref{fig:PD} (c), and (d), we see that, while the match between the two methods of evaluating $G_s(r,t)$ is not so good for intermediate times for both the confinements, match gets better for IWM as time progresses.

\section{Distinct part of VHCF at $T=0.030$}
 
\begin{figure}[h]
\begin{center}
      \includegraphics[width=15cm,keepaspectratio]{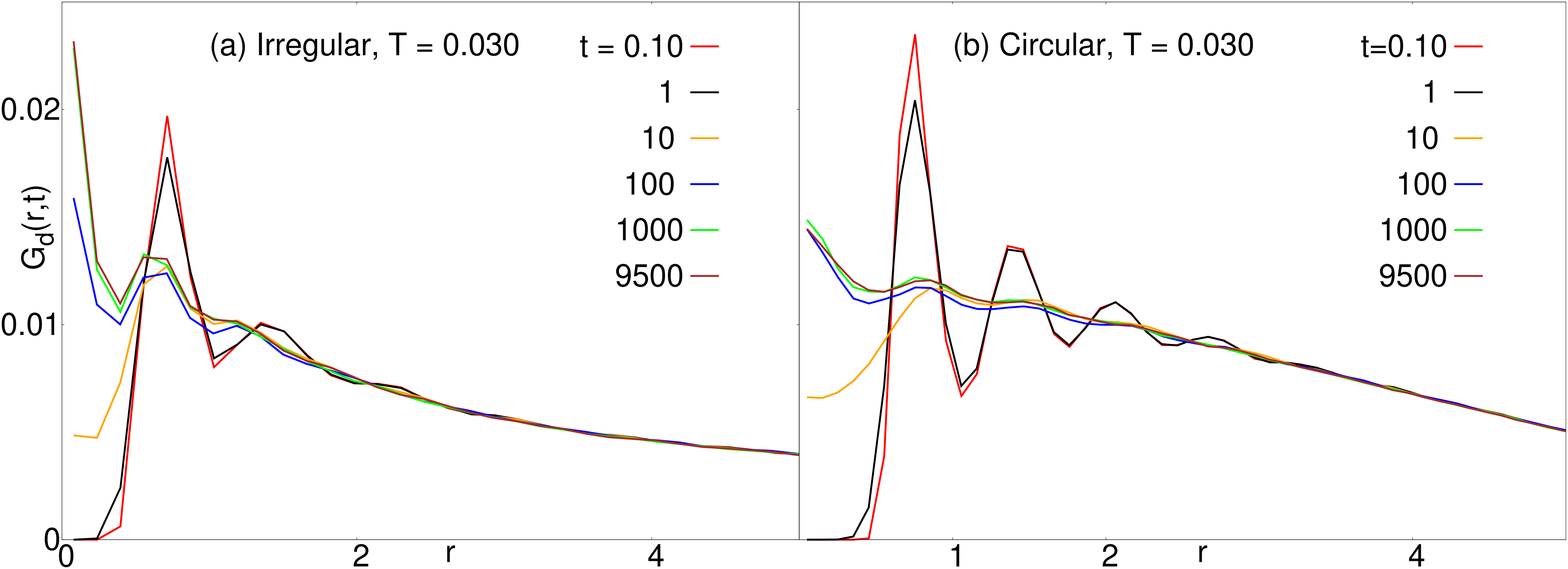} 
      \caption{$G_d(r,t)$ with $r$ for several $t$ in the `liquid', depicting the temporal relaxation in an (a) IWM and (b) CWM. The `correlation-hole' in $G_d(r \approx 0,t)$ fills up very quickly for both confinements.}
      \label{fig:DVH_03}
\end{center}
   \end{figure}

We show the distinct part of VHCF, $G_d(r,t)$, at $T=0.030$ in Fig.~\ref{fig:DVH_03}(a, b) for the IWM and CWM respectively. As expected in a liquid, the large $r$ Bragg-peaks are already washed out at smallest $t$. Its relaxation with time smoothens out even the weak modulations leaving only the strong peak at $r=0$.

\section{Slow and fast particles in a CWM}

\begin{figure}[h]
\begin{center}
      \includegraphics[width=7.0cm,keepaspectratio]{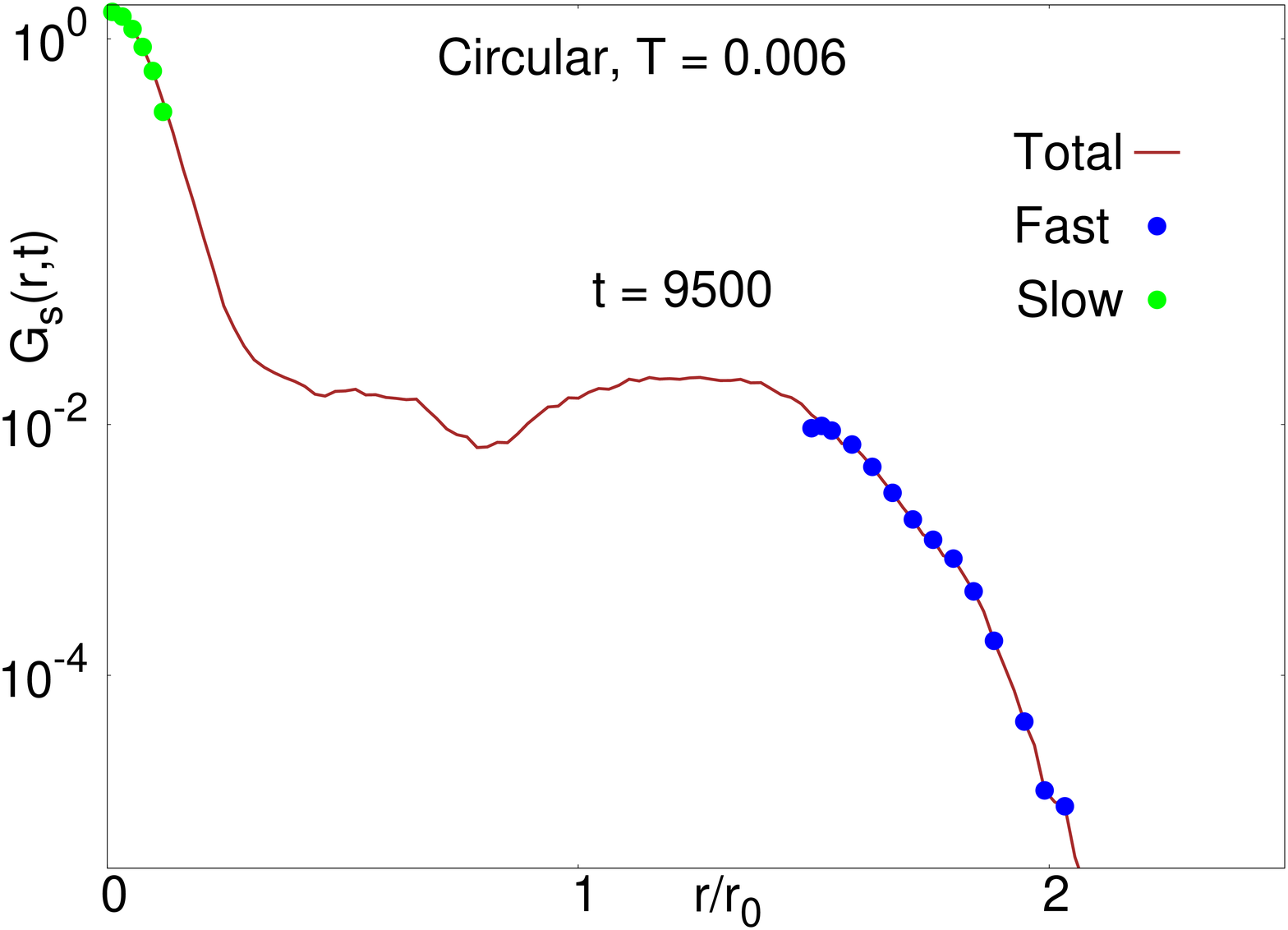} 
      \caption{Illustration of the contribution of slow (green) and fast (blue) particles to $G_s(r,t)$ for CWM. The contribution of slow and fast particles is consistent with the description given in the main text.}
      \label{fig:SlowFast}
\end{center}
   \end{figure}

In order to illustrate the complementary role of the fast and slow particles at low $T$, we identified in the main text (inset of Fig.~3(a)) the sections of a single $G_s(r,t)$ contributed by them individually for an IWM at intermediate $t$. Their independent roles can also be seen in circular trap and at large $t$, which we demonstrate here in Fig.~\ref{fig:SlowFast}.

\begin{thebibliography}{99}
\bibitem{FN0} Our scaled energy and time unit takes the value $1.7$ meV and $376$ fs, for electrons in circular quantum dot. Here, bare parameters are taken for GaAs heterostructures ($q = -e$, $m = 0.067 m_e$, the mass of an electron, $\hbar\omega_0 = 1$ meV, $\epsilon = 13 $) \cite{parameters}
\bibitem{FN1} Though the classification of temporal regimes holds for all realizations of irregularity (combination of $\{\lambda,\gamma\}$), the width of each regime show statistical fluctuations over realizations. Thus, Figs. 1, 2(a, b) and 3 display results for a specific $\lambda=0.635$ and $\gamma=0.2$.
\bibitem{FN2} The coalescing azimuthal motion in the CWM at large $t$ is different from a rigid rotation, which are easily tracked in the periodic modulation of the mean-square displacemt, and subtracted from all our results.
\bibitem{FN3} We obtain $P(D)$ using MSD, $\langle {r_i}^2(t)\rangle = 4D_it$ for all particles $i$. Also $P(D) \equiv P(\langle r^2 \rangle)$ for a fixed $t$ and the equivalence was used for convenience.
\bibitem{FN4} Note that $\lim_{t\rightarrow 0} G_d(r,t)\rightarrow g(r)$, where $g(r)$ is the pair correlation function.
\end{thebibliography}
\end{document}